\documentclass{nature}
\usepackage[switch]{lineno}

\usepackage{graphicx}
\usepackage{xcolor} 

\usepackage{bm}
\usepackage{amsmath,amsthm,amsfonts,amssymb,amscd}
\usepackage{threeparttable}
\usepackage{parskip}

\let\oldequation\equation
\let\oldendequation\endequation

\renewenvironment{equation}{\linenomathNonumbers\oldequation}{\oldendequation\endlinenomath}

\title{Evidence for baryon acoustic oscillations from galaxy-ellipticity correlations}

\author{Kun Xu$^{1,3}$, Y. P. Jing$^{1,2}$, Gong-Bo Zhao$^{4,5,6}$ \& Antonio J. Cuesta$^7$}

\begin{document}

\maketitle

\begin{affiliations}
 \item Department of Astronomy, School of Physics and Astronomy, Shanghai Jiao Tong University, Shanghai, 200240, P.R.China
 \item Tsung-Dao Lee Institute, and Shanghai Key Laboratory for Particle Physics and Cosmology, Shanghai Jiao Tong University, Shanghai, 200240, P.R.China
 \item Institute for Computational Cosmology, Durham University, South Road, Durham DH1 3LE, UK
 \item National Astronomy Observatories, Chinese Academy of Sciences, Beijing, 100101, P.R.China
 \item University of Chinese Academy of Sciences, Beijing, 100049, P.R.China
 \item Institute for Frontiers in Astronomy and Astrophysics, Beijing Normal University, Beijing, 102206, P.R.China
 \item Departamento de F\'isica, Universidad de C\'ordoba,
C\'ordoba, E-14071, Spain
\end{affiliations}

\begin{abstract}
The Baryon Acoustic Oscillations (BAO) feature in the clustering of galaxies or quasars provides a ``standard ruler" for distance measurements in cosmology. In this work, we report a $2\sim3\sigma$ signal of the BAO dip feature in the galaxy density-ellipticity (GI) cross-correlation functions using the spectroscopic sample of the Baryon Oscillation Spectroscopic Survey (BOSS) CMASS, combined with the deep DESI Legacy Imaging Surveys for precise galaxy shape measurements. We measure the GI correlation functions and model them using the linear alignment model. We constrain the distance $D_V/r_{\mathrm{d}}$ to redshift $0.57$ to a precision of $3\sim5\%$, depending on the details of modeling. The GI measurement reduces the uncertainty of distance measurement by $\sim10\%$ on top of that derived from the galaxy-galaxy (GG) correlation. More importantly, for future large and deep galaxy surveys, the independent GI measurements can help sort out the systematics in the BAO studies.

\end{abstract}

Measuring the expansion history of the Universe is one of the key goals in cosmology. The best constraints now are from the measurements of the distance-redshift relation over a wide range of redshifts\cite{2013PhR...530...87W}. The Baryon Acoustic Oscillations (BAO) feature in the clustering of galaxies is a ``standard ruler" for robust distance measurements\cite{1970ApJ...162..815P,1998ApJ...496..605E}. BAO arise from tight coupling of photons and baryons in the early Universe. Sound waves travel through this medium and give rise to a characteristic scale in the density perturbations, corresponding to the propagation distance of the waves before the recombination. With large galaxy surveys, using BAO, distances have been measured to percent level at various redshifts\cite{2005ApJ...633..560E,2010MNRAS.401.2148P,2017MNRAS.470.2617A,2018MNRAS.473.4773A,2020ApJ...901..153D}.

In the aforementioned studies, only the spatial distributions of galaxies are used while the shapes or orientations of galaxies are ignored. The intrinsic alignment (IA) of galaxies is usually treated as a contaminant in weak lensing analysis\cite{2000MNRAS.319..649H,2000ApJ...545..561C}. However, as we will demonstrate using actual observations in this work, IA is actually a promising cosmological probe and contains valuable information. The galaxy-ellipticity (GI) and ellipticity-ellipticity (II) intrinsic alignments were first detected in Luminous Red Galaxies (LRG)\cite{2007MNRAS.381.1197H, 2009ApJ...694..214O,2023arXiv230204230X}. Studies showed that the IA of galaxies can be related to the gravitational tidal field using the linear alignment (LA) model\cite{2001ApJ...559..552C,2004PhRvD..70f3526H,2011JCAP...05..010B}. According to the LA model, a BAO feature in both GI and II correlations shows up as a dip rather than a peak\cite{2013JCAP...12..029C,2020MNRAS.493L.124O,2023OJAp....6E..19V} as seen in the GG correlations. Furthermore, the entire 2D and anisotropic pattern of GI and II correlations may provide additional information to the BAO. The results were tested and confirmed in the $N$-body simulations\cite{2020MNRAS.494..694O}. Taruya \& Okumura\cite{2020ApJ...891L..42T} performed a forecast of cosmological constraints for IA statistics using the LA model, and found that IA can provide a similar level of constraints on cosmological parameters as the galaxy spatial distributions.

In this Letter, we report a first measurement of BAO feature using IA statistics, namely GI, and confirm that IA can provide additional information to galaxy-galaxy (GG) correlations. In addition to reducing statistical uncertainties of the distance measurements, GI can also provide a test of systematics when compared with the BAO measurements from GG. Details of this analysis including the datasets, measurements and modeling are presented in {\textbf{Methods}}. Our fiducial results, as shown in the main text, are derived using data vectors in the range of 50-200 $h^{-1}{\rm{Mpc}}$ with a polynomial marginalised over. This range is chosen to ensure that the polynomial captures the entire shape and avoid the turnover around $20\,h^{-1}{\rm{Mpc}}$. To be complete, we also show results with other choices of the fitting range and with or without marginalising over the polynomial in Supplementary Figure 4, 5 and Supplementary Table 1. We observe a $2-3\sigma$ signal of BAO in all of these variations.

The measurements and modeling of the isotropic GI and GG correlation functions are shown in Fig. 1. The GI measurements show an apparent dip around the BAO scale at $\sim 100\ h^{-1}\mathrm{Mpc}$ from both the pre- (panel a) and post-reconstructed samples (panel b). The fitting results are reasonable on all scales, namely, the reduced $\chi^2$ is $0.97$ for pre-reconstruction and $0.87$ for post-reconstruction. To show the significance of the BAO detection, we display the $\Delta\chi^2\equiv\chi^2-\chi^2_{\rm{min}}$ surfaces in Fig. 2 (see also {\textbf{Supplementary Figure 1}}), where $\chi_{\rm{min}}^2$ is the $\chi^2$ for the best-fitting model. We compare $\Delta\chi^2$ for the no-wiggle model with the BAO model and find a $3\sigma$ detection of the BAO feature in both the pre- and post reconstructed samples.

The constraint on $\alpha$, which represents the deviation from the fiducial cosmology (see {\textbf{Methods}}), is $1.050^{+0.030}_{-0.028}$ and $1.057^{+0.035}_{-0.036}$ using the pre- and post-reconstructed samples, respectively. Both results are in good agreement (within 2$\sigma$) with the fiducial Planck18\cite{2020A&A...641A...6P} results (TT,TE,EE+lowE+lensing+ BAO), which assumes a $\Lambda$CDM cosmology. We find that the constraint is not improved after reconstruction for the GI correlations, which may be due to the fact that we only reconstruct the density field but keep the shape field unchanged. The result may be further improved in principle, if the shape field is also reconstructed, which is left for a future study.

The post-reconstructed GG (panel c) and combined (panel d) measurements and modeling are also shown in Fig. 1. The GG correlation alone measures $\alpha$ to be $0.986^{+0.013}_{-0.013}$ and the combined GG+GI derives the constraint to be $0.997^{+0.012}_{-0.012}$, demonstrating that the GI measurement gives rise to a $\sim10\%$ improvement in terms of the uncertainty on $\alpha$. And more importantly, as we mentioned above, the next generation surveys can tighten the GG BAO constraints by a factor of 2-3\cite{2013PhR...530...87W,2016arXiv161100036D} and systematic errors become more and more important. If the GI measurements can be improved at the same level, comparisons between a sub-percent ($<0.5\%$) GG measurements and a $1\%$ GI measurements can provide a check of the systematic bias in the measurements. Our GG results are consistent with the results reported by the BOSS team\cite{2017MNRAS.470.2617A} within the $68\%$ confidence level uncertainty, although the numbers are slightly different due to a few effects, including fitting ranges, details of radial bins and error estimations.

In Fig. 3, we convert the constraints on $\alpha$ to distance $D_V/r_{\mathrm{d}}$ measured at redshift $z=0.57$ with the fiducial value $D_{V,\mathrm{fid}}(0.57)=2056.58\ \mathrm{Mpc}$, $r_{\mathrm{d,fid}}=147.21\ \mathrm{Mpc}$ and $D_{V,\mathrm{fid}}/r_{\mathrm{d,fid}}=13.97$ for Planck18\cite{2020A&A...641A...6P}. The quantity $D_V/r_{\mathrm{d}}$ is measured to be $14.67^{+0.42}_{-0.40}$, $14.77^{+0.49}_{-0.50}$, $13.77^{+0.18}_{-0.19}$ and $13.91^{+0.17}_{-0.17}$ using the pre-reconstructed GI, post-reconstructed GI, GG and GI+GG, respectively. All these results are consistent with Planck18 within $2\sigma$ level.

In the work, we obtain a $2\sim3\sigma$ measurement of the BAO dip feature in galaxy-ellipticity correlations, although the constraints on the distance from GI are only around $1/3$ of that from GG, much weaker than predicted in Taruya \& Okumura\cite{2020ApJ...891L..42T} using the linear alignment model. The reason may be that the galaxy-halo misalignment\cite{2009ApJ...694L..83O} can reduce the intrinsic alignment signals and weaken the BAO constraints, which may not be considered appropriately by Taruya \& Okumura\cite{2020ApJ...891L..42T}. According to Okumura \& Jing\cite{2009ApJ...694L..83O}, taking into account the misalignment, the GI signals can be reduced by 2-3, which is consistent with our results. Moreover, since realistic mock catalogs for galaxy shapes are unavailable yet, the covariance matrices in this study are estimated using the jackknife resampling method. Employing more reliable error estimation techniques could potentially improve the accuracy of the results, and is left for a future study. Nevertheless, the results are already promising. With the next generation spectroscopic and photometric surveys including the Dark Energy Spectroscopic Instrument (DESI) \cite{2016arXiv161100036D} and the Legacy Survey of Space and
Time (LSST)\cite{2019ApJ...873..111I}, we will have larger galaxy samples and better shape measurements. We expect that the intrinsic alignment statistics can provide much tighter constraints on cosmology from BAO and other probes \cite{2023arXiv230202925K,2023ApJ...945L..30O}.

\begin{methods}
\subsection{Statistics of the intrinsic alignment}

The shape of galaxies can be characterized by a two-component ellipticity, which is defined as follows:
\begin{equation}
    \gamma_{(+,\times)}=\frac{1-q^2}{1+q^2}(\cos{(2\theta)},\sin{(2\theta)})\,\,,\label{eq:1}
\end{equation}
where $q$ represents the minor-to-major axial ratio of the projected shape, and $\theta$ denotes the angle between the major axis projected onto the celestial sphere and the projected separation vector pointing towards a specific object. 

The GI correlations, denoted as the cross-correlation functions between density and ellipticity fields, can be expressed as \cite{2004PhRvD..70f3526H,2020MNRAS.493L.124O}
\begin{equation}
    \xi_{gi}(\bm{r})=\langle[1+\delta_g(\bm{x}_1)][1+\delta_g(\bm{x}_2)]\gamma_i(\bm{x}_2)\rangle\,\,,
\end{equation}
Here, $\bm{r}=\bm{x}_1-\bm{x}_2$, and $i=\{+,\times\}$.

In this work, we focus on the GI correlation $\xi_{g+}$ since the signal of $\xi_{g\times}$ vanishes due to parity considerations. It is worth noting that the IA statistics exhibit anisotropy even in real space due to the utilization of projected shapes of galaxies, and the presence of redshift space distortion (RSD)\cite{1987MNRAS.227....1K} can introduce additional anisotropies in $\xi_{g+}(\bm{r})$. Therefore, we define the multipole moments of the correlation functions as\cite{1992ApJ...385L...5H}
\begin{equation}
    X_{\ell}(r)=\frac{2\ell+1}{2}\int_{-1}^1d\mu X(\bm{r})\mathcal{P}_{\ell}(\mu)\,\,,
\end{equation}
Here, $X$ represents one of the correlation functions, $\mathcal{P}_{\ell}$ denotes the Legendre polynomials, and $\mu$ corresponds to the directional cosine between $\bm{r}$ and the line-of-sight direction.\\

\subsection{The linear alignment model}

On large scales, the linear alignment (LA) model is frequently employed in studies of intrinsic alignments \cite{2001ApJ...559..552C,2004PhRvD..70f3526H,2020MNRAS.493L.124O}. This model assumes a linear relationship between the ellipticity fields of galaxies and halos and the gravitational tidal field.
\begin{equation}
    \gamma_{(+,\times)}(\bm{x})=-\frac{C_1}{4\pi G}(\nabla^2_x-\nabla^2_y,2\nabla_x\nabla_y)\Psi_p(\bm{x})\,\,,\label{eq:5}
\end{equation}
where $\Psi_p$ represents the gravitational potential, $G$ denotes the gravitational constant, and $C_1$ characterizes the strength of IA. Although the observed ellipticity field is density-weighted, namely $[1+\delta_g(\bm{x})]\gamma_{+,\times}(\bm{x})$, the term $\delta_g(\bm{x})\gamma_{+,\times}(\bm{x})$ is sub-dominant on large scales\cite{2020MNRAS.493L.124O} due to $\delta_g(\bm{x})\ll1$, and it can be neglected at the BAO scale. In the Fourier space, Equation \ref{eq:5} can be expressed as:
\begin{equation}
    \gamma_{+,\times}(\bm{k})=-\widetilde{C}_1\frac{(k_x^2-k_y^2,2k_xk_y)}{k^2}\delta(\bm{k})\,\,,
\end{equation}
Here, $\widetilde{C}_1(a)\equiv a^2C_1\bar{\rho}(a)/\bar{D}(a)$, where $\bar{\rho}$ represents the mean mass density of the Universe, $\bar{D}(a)\varpropto D(a)/a$, and $D(a)$ corresponds to the linear growth factor, with $a$ denoting the scale factor.

Then, $\xi_{g+}(\bm{r})$ can be represented by the matter power spectrum $P_{\delta\delta}$\cite{2019PhRvD.100j3507O,2020MNRAS.493L.124O}:
\begin{equation}
    \xi_{g+}(\bm{r})=\widetilde{C}_1b_g(1-\mu^2)\int_0^{\infty}\frac{k^2dk}{2\pi^2}P_{\delta\delta}(k)j_{2}(kr)\,\,,
\end{equation}
where $b_g$ is the linear galaxy bias and $j_2$ is the second-order spherical Bessel function.

The RSD effect \cite{1987MNRAS.227....1K} can also be considered in $\xi_{g+}(\bm{r})$ at large scales\cite{2020MNRAS.493L.124O}. However, in this work, we do not consider the RSD effect and only focus on the \textit{monopole} component of $\xi_{g+}(\bm{r})$ given the sensitivity of current data.
\begin{equation}
    \xi_{g+,0}(r)=\frac{2}{3}\widetilde{C}_1b_g\int_0^{\infty}\frac{k^2dk}{2\pi^2}P_{\delta\delta}(k)j_{2}(kr)\,\,.\label{eq:8}
\end{equation}
 We plan to measure the entire 2D $\xi_{g+}(\bm{r})$ with future large galaxy surveys, which may contain much more information. To test the LA model, Okumura et al.\cite{2020MNRAS.494..694O} measured the IA statistics in N-body simulations and found that the results agree well with the predictions from the LA model on large scales. Thus, it is reasonable to use the above formula of $\xi_{g+,0}(r)$ for BAO studies.\\

\subsection{Fitting the BAO scale} We fit the BAO features in GG correlations following the SDSS-III BOSS DR12 analysis\cite{2017MNRAS.470.2617A,2017MNRAS.469.3762W}.

To model the BAO features in GI correlations, we adopt the similar methodology used in isotropic galaxy correlations (GG) studies\cite{2010MNRAS.401.2148P,2014MNRAS.441...24A}. In spherically averaged two-point measurements, the BAO position is fixed by the sound horizon at the baryon-drag epoch $r_{\mathrm{d}}$ and provide a measurement of\cite{2005ApJ...633..560E}
\begin{equation}
    D_V(z)\equiv[cz(1+z)^2D_A(z)^2H^{-1}(z)]^{1/3}\,\,,\label{eq:9}
\end{equation}
where $D_A(z)$ is the angular diameter distance and $H(z)$ is the Hubble parameter. The correlation functions are measured under an assumed fiducial cosmological model to convert angles and redshifts into distances. The deviation of the fiducial cosmology from the true one can be measured by comparing the BAO scale in clustering measurements to its position in a template constructed using the fiducial cosmology. The deviation is characterized by
\begin{equation}
    \alpha\equiv\frac{D_V(z)r_{\mathrm{d},\mathrm{fid}}}{D_{V,\mathrm{fid}}(z)r_\mathrm{d}}\,\,,
\end{equation}
where the subscripts "fid" denote the quantities from the fiducial cosmology.

The template of $\xi_{g+,0}$ is generated using the linear power spectrum, $P_{\mathrm{lin}}$, from the \texttt{CLASS} code\cite{2011arXiv1104.2932L}. In GG BAO peak fitting, a linear power spectrum with damped BAO is usually used to account for the non-linear effect,
\begin{equation}
    P_{\mathrm{damp}}(k)=P_{\mathrm{nw}}(k)\Big[1+\Big(\frac{P_{\mathrm{lin}}(k)}{P_{\mathrm{nw}}(k)}-1\Big)e^{-(1/2)k^2\Sigma^2_{\mathrm{nl}}}\Big]\,\,,
\end{equation}
where $P_{\mathrm{nw}}$ is the fitting formula of the no-wiggle power spectrum\cite{1998ApJ...496..605E} and $\Sigma_{\mathrm{nl}}$ is the damping scale. In this analysis, we set $\Sigma_{\mathrm{nl}}=0$ as our fiducial model for GI, and we also show the results with $\Sigma_{\mathrm{nl}}$ as a free parameter in {\textbf{Supplementary Table 1}}.

Using the template, our model for GI correlation is given by
\begin{equation}
    \xi_{g+,0}(s)=B\int_0^{\infty}\frac{k^2dk}{2\pi^2}P_{\mathrm{lin}}(k)j_{2}(\alpha ks)\,\,,\label{eq:12}
\end{equation}
where $B$ accounts for all the effects that only affect the amplitude of the correlation such as IA strength, galaxy bias and shape responsivity (see Equation 16). As in the GG analysis, we further add an additional polynomial in our model to marginalize over the broad band shape:
\begin{equation}
    \xi_{g+,0}^{{\rm{mod}}}(s)=\xi_{g+,0}(s)+\frac{a1}{s^2}+\frac{a2}{s}+a3\,\,.
\end{equation}

Thus, with the observed GI correlation $\xi^{\mathrm{obs}}_{g+,0}(s)$ and the covariance matrix $\mathbf{C}$, we can assume a likelihood function $\mathcal{L}\varpropto\exp{(-\chi^2/2)}$, with
\begin{equation}
    \chi^2=\frac{N_{\mathrm{JK}}-N_{\mathrm{bin}}-2}{N_{\mathrm{JK}}-1}\sum_{i,j}\big[\xi_i^{\mathrm{mod}}-\xi_i^{\mathrm{obs}}\big]\mathbf{C}^{-1}_{ij}\big[\xi_j^{\mathrm{mod}}-\xi_j^{\mathrm{obs}}\big]\,\,,
\end{equation}
where $\mathbf{C}^{-1}$ is the inverse of $\mathbf{C}$, $i,j$ indicate the data points at different radial bins, $N_\mathrm{JK}$ and $N_{\mathrm{bin}}$ are the total number of sub-samples and radial bins, and $(N_{\mathrm{JK}}-N_{\mathrm{bin}}-2)/(N_{\mathrm{JK}}-1)$ is the Hartlap correction factor\cite{2007A&A...464..399H} to get the unbiased covariance matrix. The covariance matrices are estimated using the jackknife resampling from the observation data:
\begin{equation}
    \mathbf{C}_{ij}=\frac{N_{\mathrm{JK}}-1}{N_\mathrm{JK}}\sum_{n=1}^{N_{\mathrm{JK}}}(\xi_i^n-\bar{\xi}_i)(\xi_j^n-\bar{\xi}_j)\,\,,
\end{equation}
where $\xi_i^{n}$ is the measurement in the $n$th sub-sample at the $i$th radial bin and $\bar{\xi}_i$ is the mean jackknife correlation function at the $i$th radial bin. We use the Markov Chain Monte Carlo (MCMC) sampler \texttt{emcee} \cite{2013PASP..125..306F} to perform a maximum likelihood analysis. $N_{\rm{JK}}$ is chosen by gradually increasing it until the constrains on $\alpha$ are stable ({\textbf{Supplementary Table 1}}). \\

\subsection{Sample selection}
We use the data from the Baryon Oscillation Spectroscopic Survey (BOSS) CMASS DR12 sample\cite{2015ApJS..219...12A,2016MNRAS.455.1553R,2017MNRAS.470.2617A}. The CMASS sample covers an effective area of $9329$ deg$^2$ and provides spectra for over $0.8$ million galaxies. Galaxies are selected with a number of magnitude and colour cuts to get an approximately constant stellar mass. We use the \texttt{CMASSLOWZTOT} LSS catalog in BOSS DR12 and adopt a reshift cut of $0.43<z<0.70$ to select the CMASS sample with an effective redshift $z_{\mathrm{eff}}=0.57$.

Reconstruction methods can improve the significance of the detection of the BAO feature, and reduce the uncertainty in BAO scale measurements, by correcting for the density field smoothing effect associated to large-scale bulk flows\cite{2007ApJ...664..675E,2009PhRvD..79f3523P,2012MNRAS.427.2132P}. We also use the post-reconstructed catalogs from BOSS DR12 and we refer the details of the reconstruction methods to their papers\cite{2012MNRAS.427.2132P,2017MNRAS.470.2617A}.

To get high quality images for the CMASS galaxies, we cross match them with the DESI Legacy Imaging Surveys DR9 data\cite{2019AJ....157..168D} (https://www.legacysurvey.org/dr9/files/\#survey-dr9-region-specobj-dr16-fits), which covers the full CMASS footprint and contains all the CMASS sources. The Legacy Surveys can reach a $r$ band PSF depth fainter than $23.5$ mag, which is 2-3 deeper than the SDSS photometry survey used for target selection and is more than adequate to study the orientations of the massive CMASS galaxies. The Legacy Surveys images are processed using \texttt{Tractor}\cite{2016ascl.soft04008L}, a forward-modeling approach to perform source extraction on pixel-level data. We use \texttt{shape\_e1} and \texttt{shape\_e2} in the Legacy Surveys DR9 catalogs (https://www.legacysurvey.org/ dr9/catalogs/\#ellipticities) as the shape measurements for each CMASS galaxy. These two quantities are then converted to the ellipticity defined in Equation \ref{eq:1}. Following Okumura \& Jing\cite{2009ApJ...694L..83O}, we assume that all the galaxies have $q=0$, which is equivalent to assuming that a galaxy is a line along its major axis. This assumption only affects the amplitude of the GI correlations, and the measurements of the position angles (PA) are more accurate than the whole galaxy shapes.

The whole CMASS sample is used to trace the density field. While for the tracers of the ellipticity field, we further select galaxies with S\'ersic index\cite{1963BAAA....6...41S} $n>2$ , since only elliptical galaxies show strong shape alignments, and $q<0.8$ for reliable PA measurements. In principle, we should also exclude satellite galaxies. However, since selecting centrals in redshift space is arbitrary and most of the CMASS LRGs are already centrals, we do not consider the central-satellite separation. Selections using $n$ and $q$ remove nearly half (from $816,779$ to $425,823$) of the CMASS galaxies. We show the results using the whole sample in {\textbf{Supplementary Table 1}} and confirm that the morphology and $q$ selection can really improve the measurements and tighten the constraints.\\

\subsection{Estimators}
To estimate the GI correlations, we generate two random samples $R_s$ and $R$ for the tracers of ellipticity and density fields respectively. Following Reid et. al.\cite{2016MNRAS.455.1553R}, redshifts are assigned to randoms to make sure that the galaxy and random catalogues have exactly the same redshift distribution. We adopt the generalized Landy–Szalay estimator \cite{1993ApJ...412...64L,2006MNRAS.367..611M}
\begin{equation}
    \xi_{g+,0}(s) = \frac{S_{+}(D-R)}{R_{s}R}\,\,,
\end{equation}
where $R_{s}R$ is the normalized random-random pairs. $S_+D$ is the sum of the $+$ component of ellipticity in all pairs:
\begin{equation}
    S_+D=\sum_{i,j}\frac{\gamma_+(j|i)}{2\mathcal{R}}\,\,,
\end{equation}\label{eq:15}
where the ellipticity of $j$th galaxy in the ellipticity tracers is defined relative to the direction to the $i$th galaxy in the density tracers, and $\mathcal{R}=1-\langle \gamma_+^2 \rangle$ is the shape responsivity \cite{2002AJ....123..583B}. $\mathcal{R}$ equals to $0.5$ under our $q=0$ assumption. $S_+R$ is calculated in a similar way using the random catalog.

We also measure the GI correlation functions for the reconstructed catalogs. The ellipticities of galaxies are assumed unchanged in the reconstruction process. The estimator becomes
\begin{equation}
    \xi_{g+,0}(s) = \frac{S_{+}(E-T)}{R_{s}R}\,\,,
\end{equation}
where $E$, $T$ represent the reconstructed data and random sample, and $R$ and $R_{s}$ are the original random samples. In above calculations, we adopt the FKP weights\cite{1994ApJ...426...23F} ($w_{{\rm{FKP}}}$) and weights for correcting the redshift failure ($w_{{\rm{zf}}}$), fibre collisions ($w_{{\rm{cp}}}$) and image systematics ($w_{{\rm{sys}}}$) for the density field tracers\cite{2016MNRAS.455.1553R}: $w_{\rm{tot}}=w_{{\rm{FKP}}}w_{{\rm{sys}}}(w_{{\rm{cp}}}+w_{{\rm{zf}}}-1)$, while no weight is used for the ellipticity field tracers. \\

\subsection{Measurements and modeling}
We measure $\xi_{g+,0}(s)$ for both pre- and post-reconstruction catalogs in $50<s<200\ h^{-1}\mathrm{Mpc}$ with a bin width of $5\ h^{-1}\mathrm{Mpc}$. We calculate their covariance matrices using the jackknife resampling with $N_{\mathrm{JK}}=400$, and model the GI correlation functions using Equation \ref{eq:12} with a Planck18\cite{2020A&A...641A...6P} fiducial cosmology at $z=0.57$. In {\textbf{Supplementary Table 1}}, we show that the pre- and post-reconstruction GI results are relatively stable if $N_{\rm{JK}}\ge400$, verifying that $N_{\rm{JK}}=400$ is a reasonable choice. We measure and model the post-reconstruction isotropic GG correlation functions with the same radial bins and error estimation schedule ($N_{\rm{JK}}=400$). We also model the GG and GI correlation together with a $60\times60$ covariance matrix that includes the GG-GI cross-covariance to get the combined results.

\subsection{Data availability}
All data used in this study is publicly available. SDSS-III BOSS DR12 LSS catalog: https://data.sdss.org/sas/dr12/boss/lss/. DESI Legacy Imaging Surveys DR9 catalog: https:// www.legacysurvey.org/dr9/catalogs/.

\end{methods}

\begin{addendum}
 \item Y.P.J. is supported by NSFC (12133006, 11890691, 11621303), grant No. CMS-CSST-2021-A03,  and 111 project No. B20019. Y.P.J. gratefully acknowledge the support of the Key Laboratory for Particle Physics, Astrophysics and Cosmology, Ministry of Education. G.-B.Z. is supported by the National Key Basic Research and Development Program of China (No. 2018YFA0404503), NSFC (11925303, 11890691), science research grants from the China Manned Space Project with No. CMS-CSST-2021-B01, and the New Cornerstone Science Foundation through the XPLORER PRIZE. A.J.C. acknowledges support from the Spanish Ministry of Science, Innovation and Universities (PID2019-107844GB-C21/AEI/10.13039/501100011033), and funding from the European Union - NextGenerationEU and the Ministerio de Universidades of Spain through Plan de Recuperaci\'on, Transformaci\'on y Resiliencia. This work made use of the Gravity Supercomputer at the Department of Astronomy, Shanghai Jiao Tong University. K.X. thanks Jun Zhang for the helpful discussions.

Funding for SDSS-III has been provided by the Alfred P. Sloan
Foundation, the Participating Institutions, the National Science
Foundation, and the US Department of Energy Office of Science. The SDSS-III web site is http://www.sdss3.org/. 

The Legacy Surveys consist of three individual and complementary projects: the Dark Energy Camera Legacy Survey (DECaLS; Proposal ID \#2014B-0404; PIs: David Schlegel and Arjun Dey), the Beijing-Arizona Sky Survey (BASS; NOAO Prop. ID \#2015A-0801; PIs: Zhou Xu and Xiaohui Fan), and the Mayall z-band Legacy Survey (MzLS; Prop. ID \#2016A-0453; PI: Arjun Dey). 
 \item[Author contributions] Y.P.J. proposed the idea and supervised the work. K.X. performed the measurements and modeling. A.J.C provided tools and expertise with the public data catalogs. K.X. and G.-B. Z. wrote the draft. All co-authors contributed to the improvement of the analysis and the manuscript.

 \item[Competing Interests] The authors declare that they have no
competing financial interests.

 \item[Correspondence] Correspondence and requests for materials should be addressed to Y.P.J. (ypjing@sjtu.edu.cn).
\end{addendum}

\newpage
\begin{figure}
\centering
\includegraphics[width=1.0\textwidth]{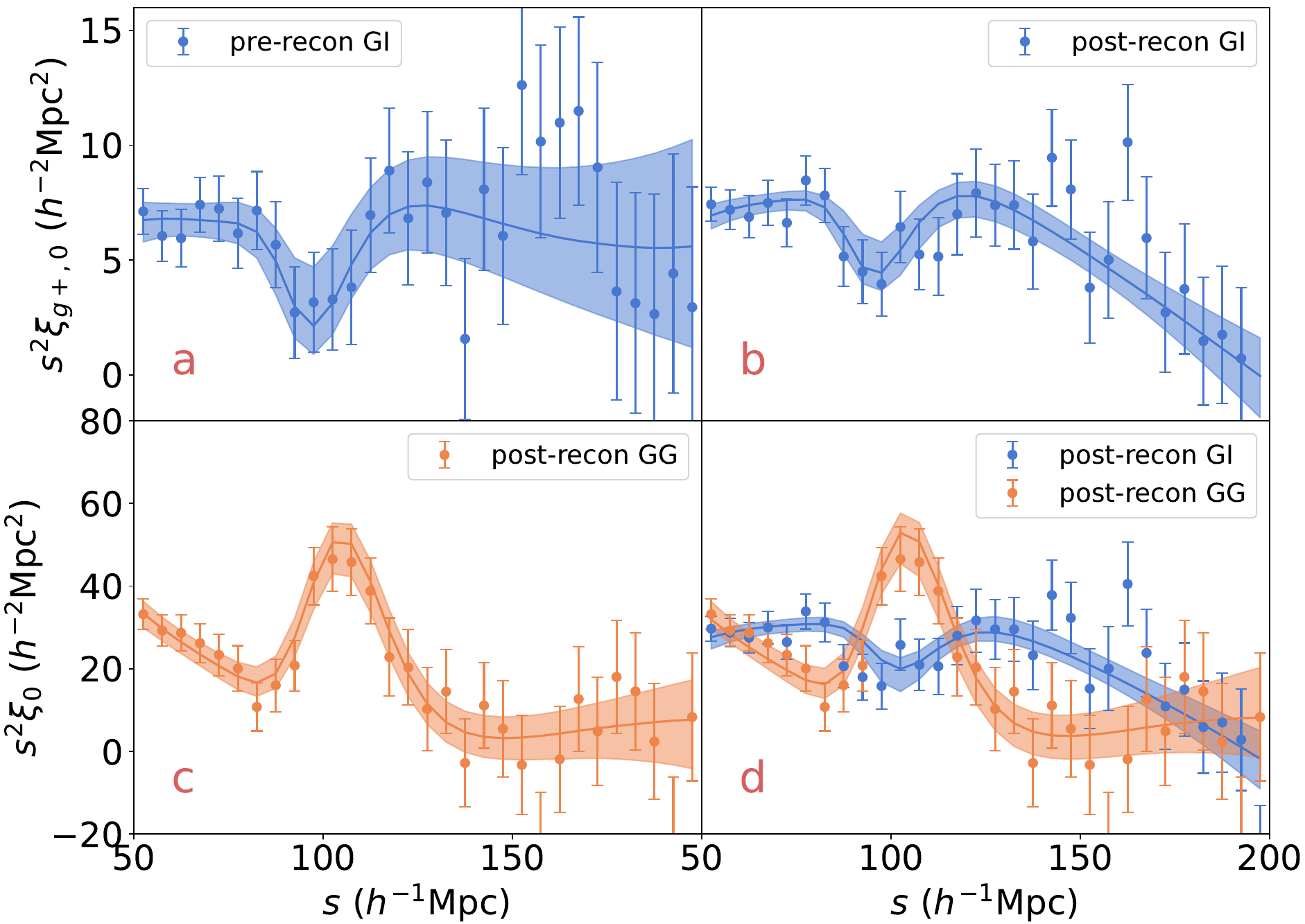}\\
\caption{Measurements and the modeling of GI and GG correlation functions. Panel a: pre-reconstruction GI correlations; Panel b: post-reconstruction GI correlations; Panel c: post-reconstruction GG correlation; Panel d: post-reconstruction combined modeling (GI multiplied by 4 for better illustration). Dots with error bars show the mean and standard error of the mean of clustering measurements. Errors are from the diagonal elements of the jackknife covariance matrices estimated using 400 subsamples. Lines with shadows are the best-fit models and $68\%$ confidence level regions derived from the marginalized posterior distributions.}\label{fig:data}
\end{figure}

\begin{figure}
\centering
\includegraphics[width=1.0\textwidth]{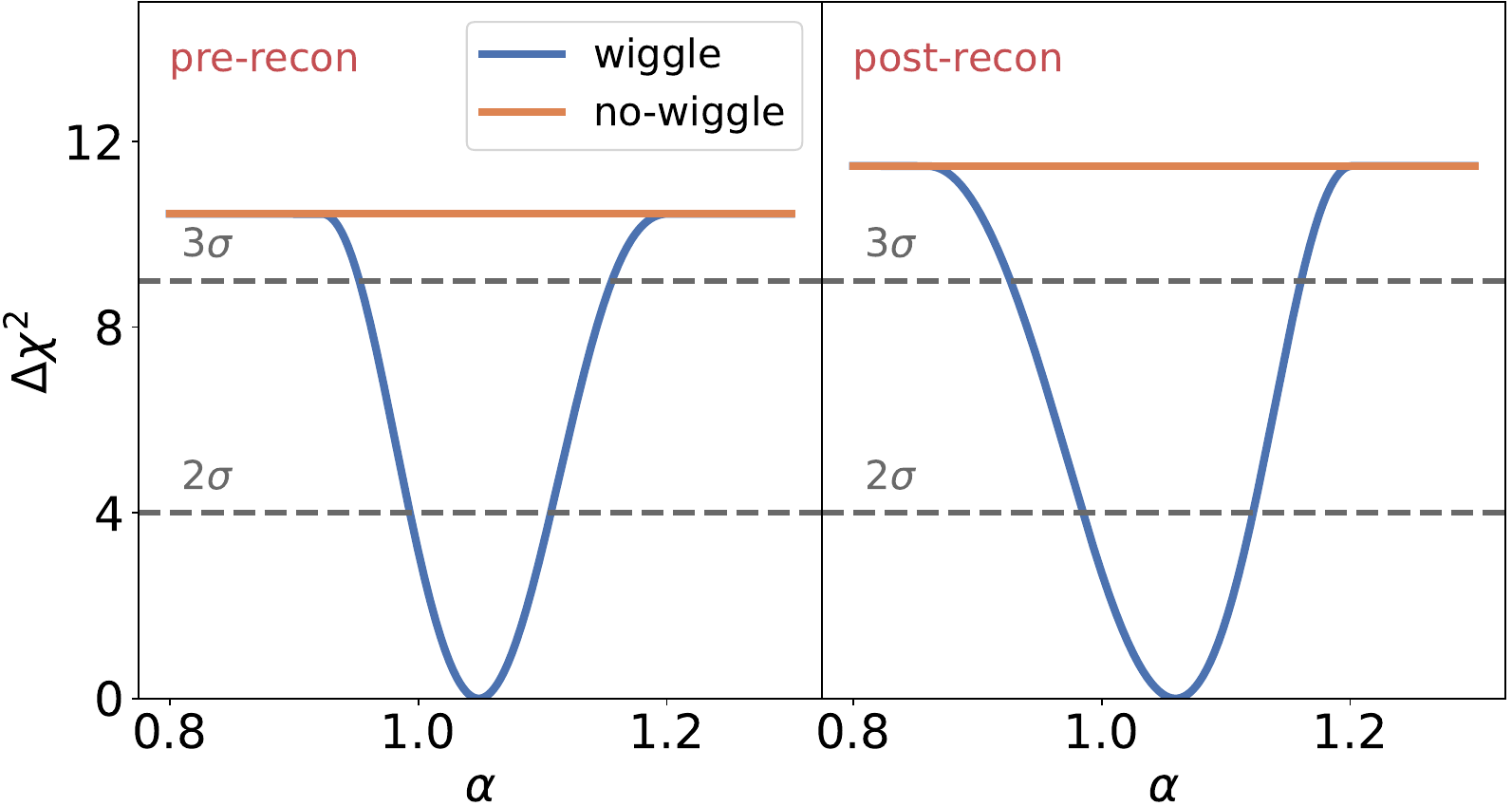}\\
\caption{Plot of $\Delta \chi^2$ versus $\alpha$ for the pre- (left) and post- (right) reconstruction GI correlations. Orange lines show the $\Delta \chi^2$ for non-BAO models and blue lines for BAO models.}
\label{fig:chi2}
\end{figure}

\begin{figure}
\centering
\includegraphics[width=1.0\textwidth]{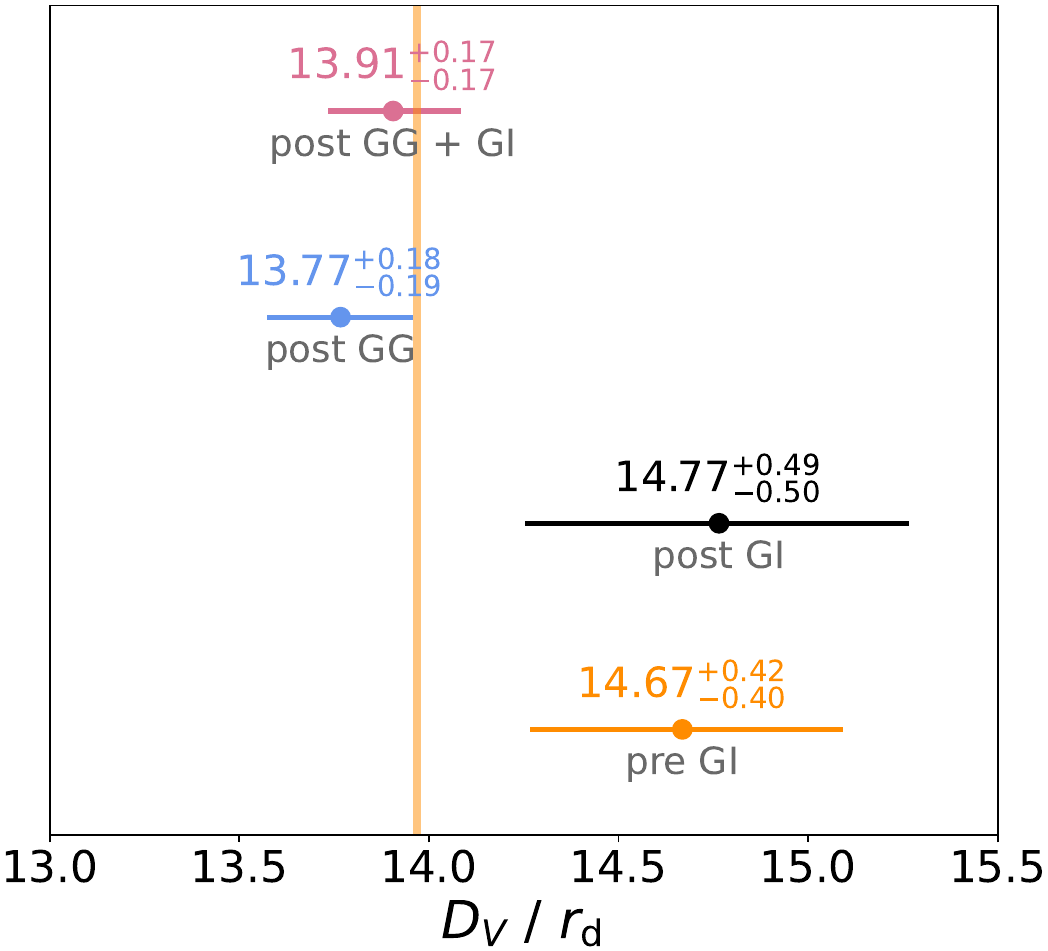}\\
\caption{Constraints of $D_V/r_{\mathrm{d}}$ from GG and GI correlation functions with $N_{\rm{JK}}=400$. A combined post-reconstructed GG+GI constraint is also provided. The central values are medians, and the errors are the 16 \& 84 percentiles. The vertical orange line shows the fiducial Planck18 results.}\label{fig:compare}
\end{figure}

\newpage
\subsection{References}
\bibliographystyle{naturemag}
\bibliography{body}

\begin{thebibliography}{10}
\expandafter\ifx\csname url\endcsname\relax
  \def\url#1{\texttt{#1}}\fi
\expandafter\ifx\csname urlprefix\endcsname\relax\def\urlprefix{URL }\fi
\providecommand{\bibinfo}[2]{#2}
\providecommand{\eprint}[2][]{\url{#2}}

\bibitem{2013PhR...530...87W}
\bibinfo{author}{{Weinberg}, D.~H.} \emph{et~al.}
\newblock \bibinfo{title}{{Observational probes of cosmic acceleration}}.
\newblock \emph{\bibinfo{journal}{Phys. Rep.}} \textbf{\bibinfo{volume}{530}},
  \bibinfo{pages}{87--255} (\bibinfo{year}{2013}).
\newblock \eprint{1201.2434}.

\bibitem{1970ApJ...162..815P}
\bibinfo{author}{{Peebles}, P.~J.~E.} \& \bibinfo{author}{{Yu}, J.~T.}
\newblock \bibinfo{title}{{Primeval Adiabatic Perturbation in an Expanding
  Universe}}.
\newblock \emph{\bibinfo{journal}{Astrophys. J.}}
  \textbf{\bibinfo{volume}{162}}, \bibinfo{pages}{815} (\bibinfo{year}{1970}).

\bibitem{1998ApJ...496..605E}
\bibinfo{author}{{Eisenstein}, D.~J.} \& \bibinfo{author}{{Hu}, W.}
\newblock \bibinfo{title}{{Baryonic Features in the Matter Transfer Function}}.
\newblock \emph{\bibinfo{journal}{Astrophys. J.}}
  \textbf{\bibinfo{volume}{496}}, \bibinfo{pages}{605--614}
  (\bibinfo{year}{1998}).
\newblock \eprint{astro-ph/9709112}.

\bibitem{2005ApJ...633..560E}
\bibinfo{author}{{Eisenstein}, D.~J.} \emph{et~al.}
\newblock \bibinfo{title}{{Detection of the Baryon Acoustic Peak in the
  Large-Scale Correlation Function of SDSS Luminous Red Galaxies}}.
\newblock \emph{\bibinfo{journal}{Astrophys. J.}}
  \textbf{\bibinfo{volume}{633}}, \bibinfo{pages}{560--574}
  (\bibinfo{year}{2005}).
\newblock \eprint{astro-ph/0501171}.

\bibitem{2010MNRAS.401.2148P}
\bibinfo{author}{{Percival}, W.~J.} \emph{et~al.}
\newblock \bibinfo{title}{{Baryon acoustic oscillations in the Sloan Digital
  Sky Survey Data Release 7 galaxy sample}}.
\newblock \emph{\bibinfo{journal}{Mon. Not. R. Astron. Soc.}}
  \textbf{\bibinfo{volume}{401}}, \bibinfo{pages}{2148--2168}
  (\bibinfo{year}{2010}).
\newblock \eprint{0907.1660}.

\bibitem{2017MNRAS.470.2617A}
\bibinfo{author}{{Alam}, S.} \emph{et~al.}
\newblock \bibinfo{title}{{The clustering of galaxies in the completed SDSS-III
  Baryon Oscillation Spectroscopic Survey: cosmological analysis of the DR12
  galaxy sample}}.
\newblock \emph{\bibinfo{journal}{Mon. Not. R. Astron. Soc.}}
  \textbf{\bibinfo{volume}{470}}, \bibinfo{pages}{2617--2652}
  (\bibinfo{year}{2017}).
\newblock \eprint{1607.03155}.

\bibitem{2018MNRAS.473.4773A}
\bibinfo{author}{{Ata}, M.} \emph{et~al.}
\newblock \bibinfo{title}{{The clustering of the SDSS-IV extended Baryon
  Oscillation Spectroscopic Survey DR14 quasar sample: first measurement of
  baryon acoustic oscillations between redshift 0.8 and 2.2}}.
\newblock \emph{\bibinfo{journal}{Mon. Not. R. Astron. Soc.}}
  \textbf{\bibinfo{volume}{473}}, \bibinfo{pages}{4773--4794}
  (\bibinfo{year}{2018}).
\newblock \eprint{1705.06373}.

\bibitem{2020ApJ...901..153D}
\bibinfo{author}{{du Mas des Bourboux}, H.} \emph{et~al.}
\newblock \bibinfo{title}{{The Completed SDSS-IV Extended Baryon Oscillation
  Spectroscopic Survey: Baryon Acoustic Oscillations with
  Ly{\ensuremath{\alpha}} Forests}}.
\newblock \emph{\bibinfo{journal}{Astrophys. J.}}
  \textbf{\bibinfo{volume}{901}}, \bibinfo{pages}{153} (\bibinfo{year}{2020}).
\newblock \eprint{2007.08995}.

\bibitem{2000MNRAS.319..649H}
\bibinfo{author}{{Heavens}, A.}, \bibinfo{author}{{Refregier}, A.} \&
  \bibinfo{author}{{Heymans}, C.}
\newblock \bibinfo{title}{{Intrinsic correlation of galaxy shapes: implications
  for weak lensing measurements}}.
\newblock \emph{\bibinfo{journal}{Mon. Not. R. Astron. Soc.}}
  \textbf{\bibinfo{volume}{319}}, \bibinfo{pages}{649--656}
  (\bibinfo{year}{2000}).
\newblock \eprint{astro-ph/0005269}.

\bibitem{2000ApJ...545..561C}
\bibinfo{author}{{Croft}, R. A.~C.} \& \bibinfo{author}{{Metzler}, C.~A.}
\newblock \bibinfo{title}{{Weak-Lensing Surveys and the Intrinsic Correlation
  of Galaxy Ellipticities}}.
\newblock \emph{\bibinfo{journal}{Astrophys. J.}}
  \textbf{\bibinfo{volume}{545}}, \bibinfo{pages}{561--571}
  (\bibinfo{year}{2000}).
\newblock \eprint{astro-ph/0005384}.

\bibitem{2007MNRAS.381.1197H}
\bibinfo{author}{{Hirata}, C.~M.} \emph{et~al.}
\newblock \bibinfo{title}{{Intrinsic galaxy alignments from the 2SLAQ and SDSS
  surveys: luminosity and redshift scalings and implications for weak lensing
  surveys}}.
\newblock \emph{\bibinfo{journal}{Mon. Not. R. Astron. Soc.}}
  \textbf{\bibinfo{volume}{381}}, \bibinfo{pages}{1197--1218}
  (\bibinfo{year}{2007}).
\newblock \eprint{astro-ph/0701671}.

\bibitem{2009ApJ...694..214O}
\bibinfo{author}{{Okumura}, T.}, \bibinfo{author}{{Jing}, Y.~P.} \&
  \bibinfo{author}{{Li}, C.}
\newblock \bibinfo{title}{{Intrinsic Ellipticity Correlation of SDSS Luminous
  Red Galaxies and Misalignment with Their Host Dark Matter Halos}}.
\newblock \emph{\bibinfo{journal}{Astrophys. J.}}
  \textbf{\bibinfo{volume}{694}}, \bibinfo{pages}{214--221}
  (\bibinfo{year}{2009}).
\newblock \eprint{0809.3790}.

\bibitem{2023arXiv230204230X}
\bibinfo{author}{{Xu}, K.}, \bibinfo{author}{{Jing}, Y.~P.} \&
  \bibinfo{author}{{Gao}, H.}
\newblock \bibinfo{title}{{Mass Dependence of Galaxy-Halo Alignment in LOWZ and
  CMASS}}.
\newblock \emph{\bibinfo{journal}{arXiv e-prints}}
  \bibinfo{pages}{arXiv:2302.04230} (\bibinfo{year}{2023}).
\newblock \eprint{2302.04230}.

\bibitem{2001ApJ...559..552C}
\bibinfo{author}{{Crittenden}, R.~G.}, \bibinfo{author}{{Natarajan}, P.},
  \bibinfo{author}{{Pen}, U.-L.} \& \bibinfo{author}{{Theuns}, T.}
\newblock \bibinfo{title}{{Spin-induced Galaxy Alignments and Their
  Implications for Weak-Lensing Measurements}}.
\newblock \emph{\bibinfo{journal}{Astrophys. J.}}
  \textbf{\bibinfo{volume}{559}}, \bibinfo{pages}{552--571}
  (\bibinfo{year}{2001}).
\newblock \eprint{astro-ph/0009052}.

\bibitem{2004PhRvD..70f3526H}
\bibinfo{author}{{Hirata}, C.~M.} \& \bibinfo{author}{{Seljak}, U.}
\newblock \bibinfo{title}{{Intrinsic alignment-lensing interference as a
  contaminant of cosmic shear}}.
\newblock \emph{\bibinfo{journal}{Phys. Rev. D}} \textbf{\bibinfo{volume}{70}},
  \bibinfo{pages}{063526} (\bibinfo{year}{2004}).
\newblock \eprint{astro-ph/0406275}.

\bibitem{2011JCAP...05..010B}
\bibinfo{author}{{Blazek}, J.}, \bibinfo{author}{{McQuinn}, M.} \&
  \bibinfo{author}{{Seljak}, U.}
\newblock \bibinfo{title}{{Testing the tidal alignment model of galaxy
  intrinsic alignment}}.
\newblock \emph{\bibinfo{journal}{J. Cosmol. Astropart. Phys.}}
  \textbf{\bibinfo{volume}{2011}}, \bibinfo{pages}{010} (\bibinfo{year}{2011}).
\newblock \eprint{1101.4017}.

\bibitem{2013JCAP...12..029C}
\bibinfo{author}{{Chisari}, N.~E.} \& \bibinfo{author}{{Dvorkin}, C.}
\newblock \bibinfo{title}{{Cosmological information in the intrinsic alignments
  of luminous red galaxies}}.
\newblock \emph{\bibinfo{journal}{J. Cosmol. Astropart. Phys.}}
  \textbf{\bibinfo{volume}{2013}}, \bibinfo{pages}{029} (\bibinfo{year}{2013}).
\newblock \eprint{1308.5972}.

\bibitem{2020MNRAS.493L.124O}
\bibinfo{author}{{Okumura}, T.} \& \bibinfo{author}{{Taruya}, A.}
\newblock \bibinfo{title}{{Anisotropies of galaxy ellipticity correlations in
  real and redshift space: angular dependence in linear tidal alignment
  model}}.
\newblock \emph{\bibinfo{journal}{Mon. Not. R. Astron. Soc.}}
  \textbf{\bibinfo{volume}{493}}, \bibinfo{pages}{L124--L128}
  (\bibinfo{year}{2020}).
\newblock \eprint{1912.04118}.

\bibitem{2023OJAp....6E..19V}
\bibinfo{author}{{van Dompseler}, D.}, \bibinfo{author}{{Georgiou}, C.} \&
  \bibinfo{author}{{Chisari}, N.~E.}
\newblock \bibinfo{title}{{The alignment of galaxies at the Baryon Acoustic
  Oscillation scale}}.
\newblock \emph{\bibinfo{journal}{The Open Journal of Astrophysics}}
  \textbf{\bibinfo{volume}{6}}, \bibinfo{pages}{19} (\bibinfo{year}{2023}).
\newblock \eprint{2301.04649}.

\bibitem{2020MNRAS.494..694O}
\bibinfo{author}{{Okumura}, T.}, \bibinfo{author}{{Taruya}, A.} \&
  \bibinfo{author}{{Nishimichi}, T.}
\newblock \bibinfo{title}{{Testing tidal alignment models for anisotropic
  correlations of halo ellipticities with N-body simulations}}.
\newblock \emph{\bibinfo{journal}{Mon. Not. R. Astron. Soc.}}
  \textbf{\bibinfo{volume}{494}}, \bibinfo{pages}{694--702}
  (\bibinfo{year}{2020}).
\newblock \eprint{2001.05302}.

\bibitem{2020ApJ...891L..42T}
\bibinfo{author}{{Taruya}, A.} \& \bibinfo{author}{{Okumura}, T.}
\newblock \bibinfo{title}{{Improving Geometric and Dynamical Constraints on
  Cosmology with Intrinsic Alignments of Galaxies}}.
\newblock \emph{\bibinfo{journal}{Astrophys. J. Lett.}}
  \textbf{\bibinfo{volume}{891}}, \bibinfo{pages}{L42} (\bibinfo{year}{2020}).
\newblock \eprint{2001.05962}.

\bibitem{2020A&A...641A...6P}
\bibinfo{author}{{Planck Collaboration}} \emph{et~al.}
\newblock \bibinfo{title}{{Planck 2018 results. VI. Cosmological parameters}}.
\newblock \emph{\bibinfo{journal}{Astron. \& Astrophys.}}
  \textbf{\bibinfo{volume}{641}}, \bibinfo{pages}{A6} (\bibinfo{year}{2020}).
\newblock \eprint{1807.06209}.

\bibitem{2016arXiv161100036D}
\bibinfo{author}{{DESI Collaboration}} \emph{et~al.}
\newblock \bibinfo{title}{{The DESI Experiment Part I: Science,Targeting, and
  Survey Design}}.
\newblock \emph{\bibinfo{journal}{arXiv e-prints}}
  \bibinfo{pages}{arXiv:1611.00036} (\bibinfo{year}{2016}).
\newblock \eprint{1611.00036}.

\bibitem{2009ApJ...694L..83O}
\bibinfo{author}{{Okumura}, T.} \& \bibinfo{author}{{Jing}, Y.~P.}
\newblock \bibinfo{title}{{The Gravitational Shear-Intrinsic Ellipticity
  Correlation Functions of Luminous Red Galaxies in Observation and in the
  {\ensuremath{\Lambda}}CDM Model}}.
\newblock \emph{\bibinfo{journal}{Astrophys. J. Lett.}}
  \textbf{\bibinfo{volume}{694}}, \bibinfo{pages}{L83--L86}
  (\bibinfo{year}{2009}).
\newblock \eprint{0812.2935}.

\bibitem{2019ApJ...873..111I}
\bibinfo{author}{{Ivezi{\'c}}, {\v{Z}}.} \emph{et~al.}
\newblock \bibinfo{title}{{LSST: From Science Drivers to Reference Design and
  Anticipated Data Products}}.
\newblock \emph{\bibinfo{journal}{Astrophys. J.}}
  \textbf{\bibinfo{volume}{873}}, \bibinfo{pages}{111} (\bibinfo{year}{2019}).
\newblock \eprint{0805.2366}.

\bibitem{2023arXiv230202925K}
\bibinfo{author}{{Kurita}, T.} \& \bibinfo{author}{{Takada}, M.}
\newblock \bibinfo{title}{{Constraints on anisotropic primordial
  non-Gaussianity from intrinsic alignments of SDSS-III BOSS galaxies}}.
\newblock \emph{\bibinfo{journal}{arXiv e-prints}}
  \bibinfo{pages}{arXiv:2302.02925} (\bibinfo{year}{2023}).
\newblock \eprint{2302.02925}.

\bibitem{2023ApJ...945L..30O}
\bibinfo{author}{{Okumura}, T.} \& \bibinfo{author}{{Taruya}, A.}
\newblock \bibinfo{title}{{First Constraints on Growth Rate from Redshift-space
  Ellipticity Correlations of SDSS Galaxies at 0.16 < z < 0.70}}.
\newblock \emph{\bibinfo{journal}{Astrophys. J. Lett.}}
  \textbf{\bibinfo{volume}{945}}, \bibinfo{pages}{L30} (\bibinfo{year}{2023}).
\newblock \eprint{2301.06273}.

\bibitem{1987MNRAS.227....1K}
\bibinfo{author}{{Kaiser}, N.}
\newblock \bibinfo{title}{{Clustering in real space and in redshift space}}.
\newblock \emph{\bibinfo{journal}{Mon. Not. R. Astron. Soc.}}
  \textbf{\bibinfo{volume}{227}}, \bibinfo{pages}{1--21}
  (\bibinfo{year}{1987}).

\bibitem{1992ApJ...385L...5H}
\bibinfo{author}{{Hamilton}, A.~J.~S.}
\newblock \bibinfo{title}{{Measuring Omega and the Real Correlation Function
  from the Redshift Correlation Function}}.
\newblock \emph{\bibinfo{journal}{Astrophys. J. Lett.}}
  \textbf{\bibinfo{volume}{385}}, \bibinfo{pages}{L5} (\bibinfo{year}{1992}).

\bibitem{2019PhRvD.100j3507O}
\bibinfo{author}{{Okumura}, T.}, \bibinfo{author}{{Taruya}, A.} \&
  \bibinfo{author}{{Nishimichi}, T.}
\newblock \bibinfo{title}{{Intrinsic alignment statistics of density and
  velocity fields at large scales: Formulation, modeling, and baryon acoustic
  oscillation features}}.
\newblock \emph{\bibinfo{journal}{Phys. Rev. D}}
  \textbf{\bibinfo{volume}{100}}, \bibinfo{pages}{103507}
  (\bibinfo{year}{2019}).
\newblock \eprint{1907.00750}.

\bibitem{2017MNRAS.469.3762W}
\bibinfo{author}{{Wang}, Y.} \emph{et~al.}
\newblock \bibinfo{title}{{The clustering of galaxies in the completed SDSS-III
  Baryon Oscillation Spectroscopic Survey: tomographic BAO analysis of DR12
  combined sample in configuration space}}.
\newblock \emph{\bibinfo{journal}{Mon. Not. R. Astron. Soc.}}
  \textbf{\bibinfo{volume}{469}}, \bibinfo{pages}{3762--3774}
  (\bibinfo{year}{2017}).
\newblock \eprint{1607.03154}.

\bibitem{2014MNRAS.441...24A}
\bibinfo{author}{{Anderson}, L.} \emph{et~al.}
\newblock \bibinfo{title}{{The clustering of galaxies in the SDSS-III Baryon
  Oscillation Spectroscopic Survey: baryon acoustic oscillations in the Data
  Releases 10 and 11 Galaxy samples}}.
\newblock \emph{\bibinfo{journal}{Mon. Not. R. Astron. Soc.}}
  \textbf{\bibinfo{volume}{441}}, \bibinfo{pages}{24--62}
  (\bibinfo{year}{2014}).
\newblock \eprint{1312.4877}.

\bibitem{2011arXiv1104.2932L}
\bibinfo{author}{{Lesgourgues}, J.}
\newblock \bibinfo{title}{{The Cosmic Linear Anisotropy Solving System (CLASS)
  I: Overview}}.
\newblock \emph{\bibinfo{journal}{arXiv e-prints}}
  \bibinfo{pages}{arXiv:1104.2932} (\bibinfo{year}{2011}).
\newblock \eprint{1104.2932}.

\bibitem{2007A&A...464..399H}
\bibinfo{author}{{Hartlap}, J.}, \bibinfo{author}{{Simon}, P.} \&
  \bibinfo{author}{{Schneider}, P.}
\newblock \bibinfo{title}{{Why your model parameter confidences might be too
  optimistic. Unbiased estimation of the inverse covariance matrix}}.
\newblock \emph{\bibinfo{journal}{Astron. \& Astrophys.}}
  \textbf{\bibinfo{volume}{464}}, \bibinfo{pages}{399--404}
  (\bibinfo{year}{2007}).
\newblock \eprint{astro-ph/0608064}.

\bibitem{2013PASP..125..306F}
\bibinfo{author}{{Foreman-Mackey}, D.}, \bibinfo{author}{{Hogg}, D.~W.},
  \bibinfo{author}{{Lang}, D.} \& \bibinfo{author}{{Goodman}, J.}
\newblock \bibinfo{title}{{emcee: The MCMC Hammer}}.
\newblock \emph{\bibinfo{journal}{Publ. Astron. Soc. Pac.}}
  \textbf{\bibinfo{volume}{125}}, \bibinfo{pages}{306} (\bibinfo{year}{2013}).
\newblock \eprint{1202.3665}.

\bibitem{2015ApJS..219...12A}
\bibinfo{author}{{Alam}, S.} \emph{et~al.}
\newblock \bibinfo{title}{{The Eleventh and Twelfth Data Releases of the Sloan
  Digital Sky Survey: Final Data from SDSS-III}}.
\newblock \emph{\bibinfo{journal}{Astrophys. J. Suppl.}}
  \textbf{\bibinfo{volume}{219}}, \bibinfo{pages}{12} (\bibinfo{year}{2015}).
\newblock \eprint{1501.00963}.

\bibitem{2016MNRAS.455.1553R}
\bibinfo{author}{{Reid}, B.} \emph{et~al.}
\newblock \bibinfo{title}{{SDSS-III Baryon Oscillation Spectroscopic Survey
  Data Release 12: galaxy target selection and large-scale structure
  catalogues}}.
\newblock \emph{\bibinfo{journal}{Mon. Not. R. Astron. Soc.}}
  \textbf{\bibinfo{volume}{455}}, \bibinfo{pages}{1553--1573}
  (\bibinfo{year}{2016}).
\newblock \eprint{1509.06529}.

\bibitem{2007ApJ...664..675E}
\bibinfo{author}{{Eisenstein}, D.~J.}, \bibinfo{author}{{Seo}, H.-J.},
  \bibinfo{author}{{Sirko}, E.} \& \bibinfo{author}{{Spergel}, D.~N.}
\newblock \bibinfo{title}{{Improving Cosmological Distance Measurements by
  Reconstruction of the Baryon Acoustic Peak}}.
\newblock \emph{\bibinfo{journal}{Astrophys. J.}}
  \textbf{\bibinfo{volume}{664}}, \bibinfo{pages}{675--679}
  (\bibinfo{year}{2007}).
\newblock \eprint{astro-ph/0604362}.

\bibitem{2009PhRvD..79f3523P}
\bibinfo{author}{{Padmanabhan}, N.}, \bibinfo{author}{{White}, M.} \&
  \bibinfo{author}{{Cohn}, J.~D.}
\newblock \bibinfo{title}{{Reconstructing baryon oscillations: A Lagrangian
  theory perspective}}.
\newblock \emph{\bibinfo{journal}{Phys. Rev. D}} \textbf{\bibinfo{volume}{79}},
  \bibinfo{pages}{063523} (\bibinfo{year}{2009}).
\newblock \eprint{0812.2905}.

\bibitem{2012MNRAS.427.2132P}
\bibinfo{author}{{Padmanabhan}, N.} \emph{et~al.}
\newblock \bibinfo{title}{{A 2 per cent distance to z = 0.35 by reconstructing
  baryon acoustic oscillations - I. Methods and application to the Sloan
  Digital Sky Survey}}.
\newblock \emph{\bibinfo{journal}{Mon. Not. R. Astron. Soc.}}
  \textbf{\bibinfo{volume}{427}}, \bibinfo{pages}{2132--2145}
  (\bibinfo{year}{2012}).
\newblock \eprint{1202.0090}.

\bibitem{2019AJ....157..168D}
\bibinfo{author}{{Dey}, A.} \emph{et~al.}
\newblock \bibinfo{title}{{Overview of the DESI Legacy Imaging Surveys}}.
\newblock \emph{\bibinfo{journal}{Astron. J.}} \textbf{\bibinfo{volume}{157}},
  \bibinfo{pages}{168} (\bibinfo{year}{2019}).
\newblock \eprint{1804.08657}.

\bibitem{2016ascl.soft04008L}
\bibinfo{author}{{Lang}, D.}, \bibinfo{author}{{Hogg}, D.~W.} \&
  \bibinfo{author}{{Mykytyn}, D.}
\newblock \bibinfo{title}{{The Tractor: Probabilistic astronomical source
  detection and measurement}}.
\newblock \bibinfo{howpublished}{Astrophysics Source Code Library, record
  ascl:1604.008} (\bibinfo{year}{2016}).
\newblock \eprint{1604.008}.

\bibitem{1963BAAA....6...41S}
\bibinfo{author}{{S{\'e}rsic}, J.~L.}
\newblock \bibinfo{title}{{Influence of the atmospheric and instrumental
  dispersion on the brightness distribution in a galaxy}}.
\newblock \emph{\bibinfo{journal}{Boletin de la Asociacion Argentina de
  Astronomia La Plata Argentina}} \textbf{\bibinfo{volume}{6}},
  \bibinfo{pages}{41--43} (\bibinfo{year}{1963}).

\bibitem{1993ApJ...412...64L}
\bibinfo{author}{{Landy}, S.~D.} \& \bibinfo{author}{{Szalay}, A.~S.}
\newblock \bibinfo{title}{{Bias and Variance of Angular Correlation
  Functions}}.
\newblock \emph{\bibinfo{journal}{Astrophys. J.}}
  \textbf{\bibinfo{volume}{412}}, \bibinfo{pages}{64} (\bibinfo{year}{1993}).

\bibitem{2006MNRAS.367..611M}
\bibinfo{author}{{Mandelbaum}, R.}, \bibinfo{author}{{Hirata}, C.~M.},
  \bibinfo{author}{{Ishak}, M.}, \bibinfo{author}{{Seljak}, U.} \&
  \bibinfo{author}{{Brinkmann}, J.}
\newblock \bibinfo{title}{{Detection of large-scale intrinsic
  ellipticity-density correlation from the Sloan Digital Sky Survey and
  implications for weak lensing surveys}}.
\newblock \emph{\bibinfo{journal}{Mon. Not. R. Astron. Soc.}}
  \textbf{\bibinfo{volume}{367}}, \bibinfo{pages}{611--626}
  (\bibinfo{year}{2006}).
\newblock \eprint{astro-ph/0509026}.

\bibitem{2002AJ....123..583B}
\bibinfo{author}{{Bernstein}, G.~M.} \& \bibinfo{author}{{Jarvis}, M.}
\newblock \bibinfo{title}{{Shapes and Shears, Stars and Smears: Optimal
  Measurements for Weak Lensing}}.
\newblock \emph{\bibinfo{journal}{Astron. J.}} \textbf{\bibinfo{volume}{123}},
  \bibinfo{pages}{583--618} (\bibinfo{year}{2002}).
\newblock \eprint{astro-ph/0107431}.

\bibitem{1994ApJ...426...23F}
\bibinfo{author}{{Feldman}, H.~A.}, \bibinfo{author}{{Kaiser}, N.} \&
  \bibinfo{author}{{Peacock}, J.~A.}
\newblock \bibinfo{title}{{Power-Spectrum Analysis of Three-dimensional
  Redshift Surveys}}.
\newblock \emph{\bibinfo{journal}{Astrophys. J.}}
  \textbf{\bibinfo{volume}{426}}, \bibinfo{pages}{23} (\bibinfo{year}{1994}).
\newblock \eprint{astro-ph/9304022}.

\end{thebibliography}

\newpage
\section*{Supplementary Information}
\setcounter{figure}{0}
\setcounter{table}{0}

\begin{figure}
\centering
\includegraphics[width=1.0\textwidth]{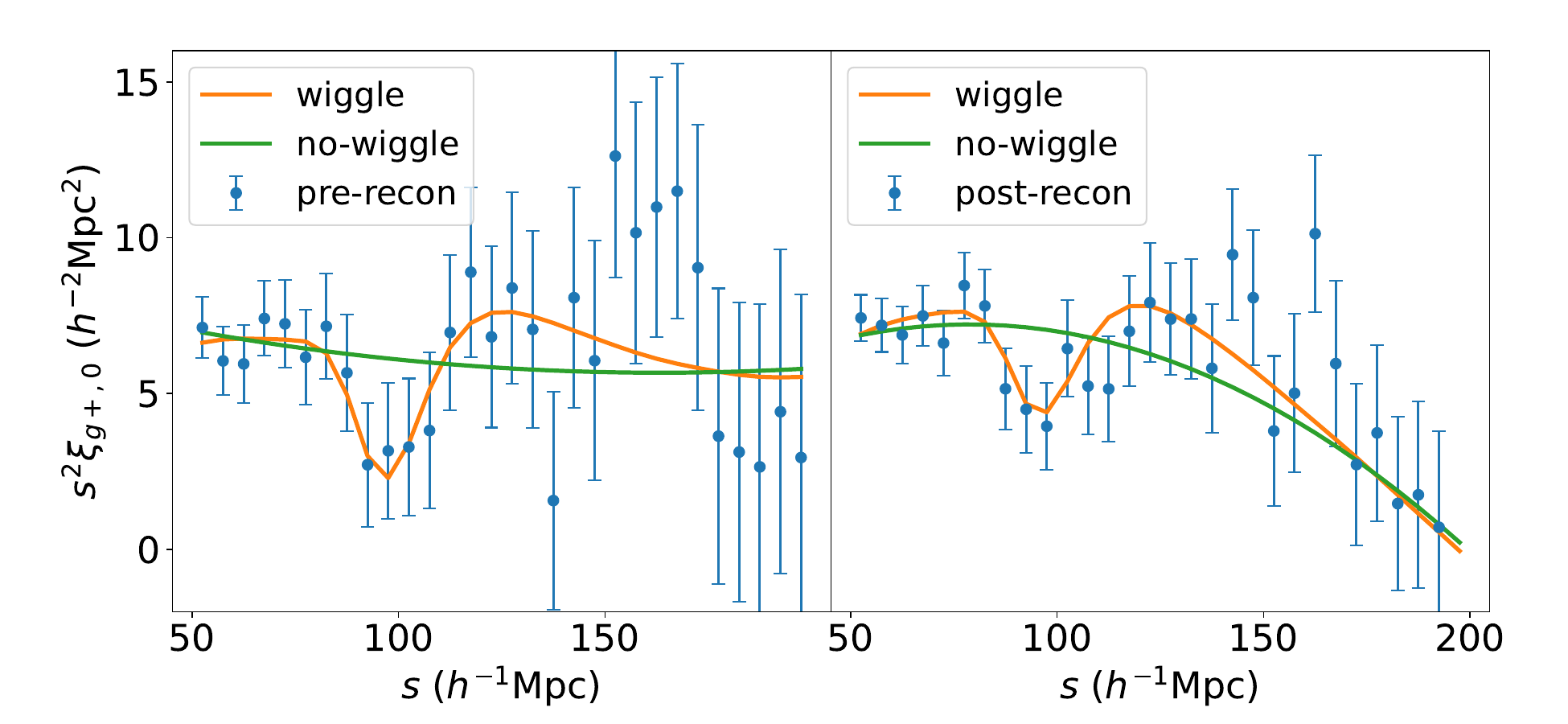}\\
\normalsize{{\textbf{Supplementary Figure 1}}: Comparison of the best-fit BAO and non-BAO models for the pre (left) and post (right) reconstruction GI correlations with fiducial models. Dots with error bars show the mean and standard error of the mean of clustering measurements. Errors are from the diagonal elements of the jackknife covariance matrices estimated using 400 subsamples.}\label{fig:fit}
\end{figure}

We compare the best-fit BAO and non-BAO models for the pre and post-reconstruction GI correlations in {\textbf{Supplementary Figure 1}}. The BAO model fits the measurements much better than the non-BAO model. 

\begin{figure}
\centering
\includegraphics[width=1.0\textwidth]{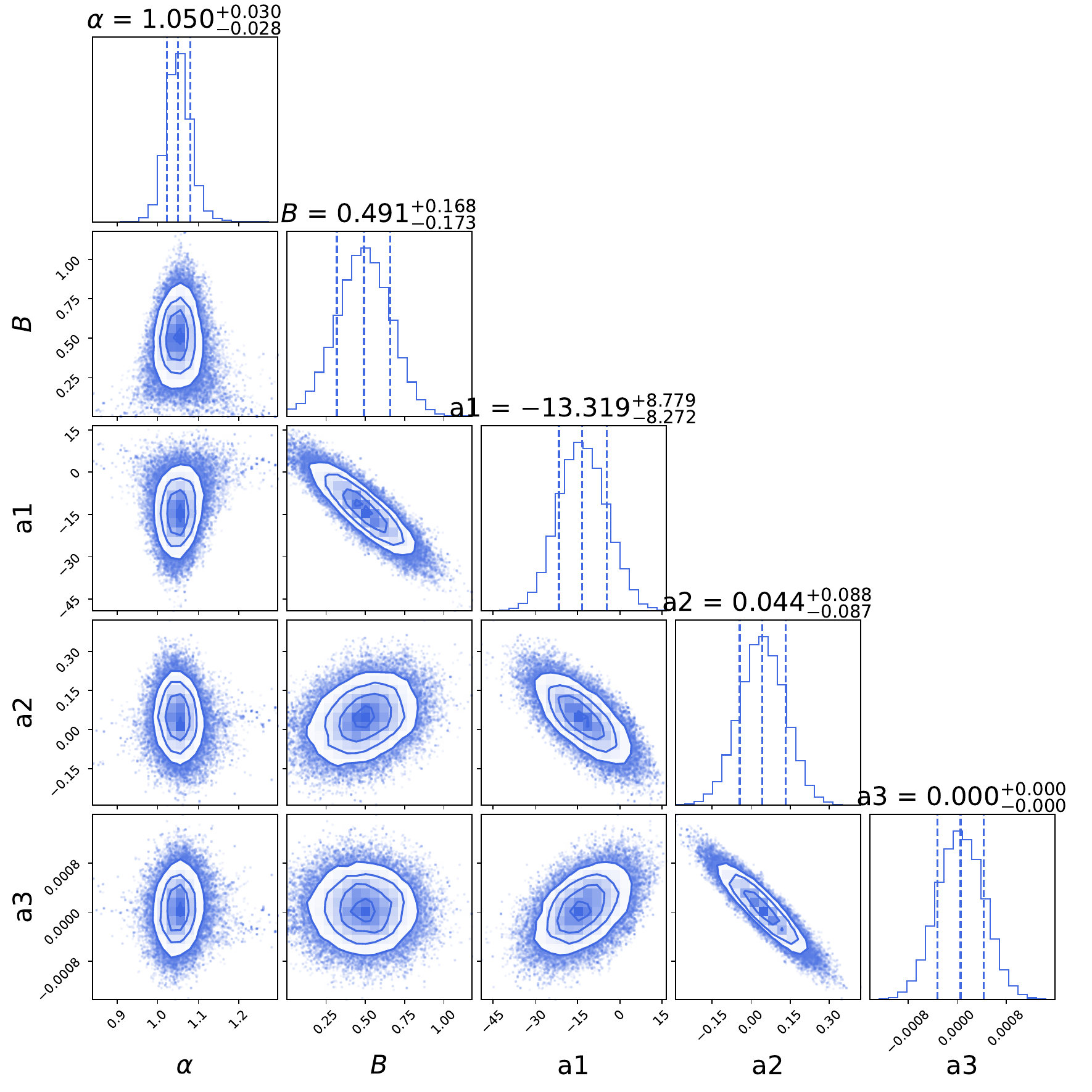}\\
\normalsize{{\textbf{Supplementary Figure 2}}: Constraints from the pre-reconstruction GI correlation functions using MCMC sampling with fiducial model. The central values are medians, and the errors are the 16 \& 84
percentiles after other parameters are marginalized over.}\label{fig:mcmc_pre}
\end{figure}

\begin{figure}
\centering
\includegraphics[width=1.0\textwidth]{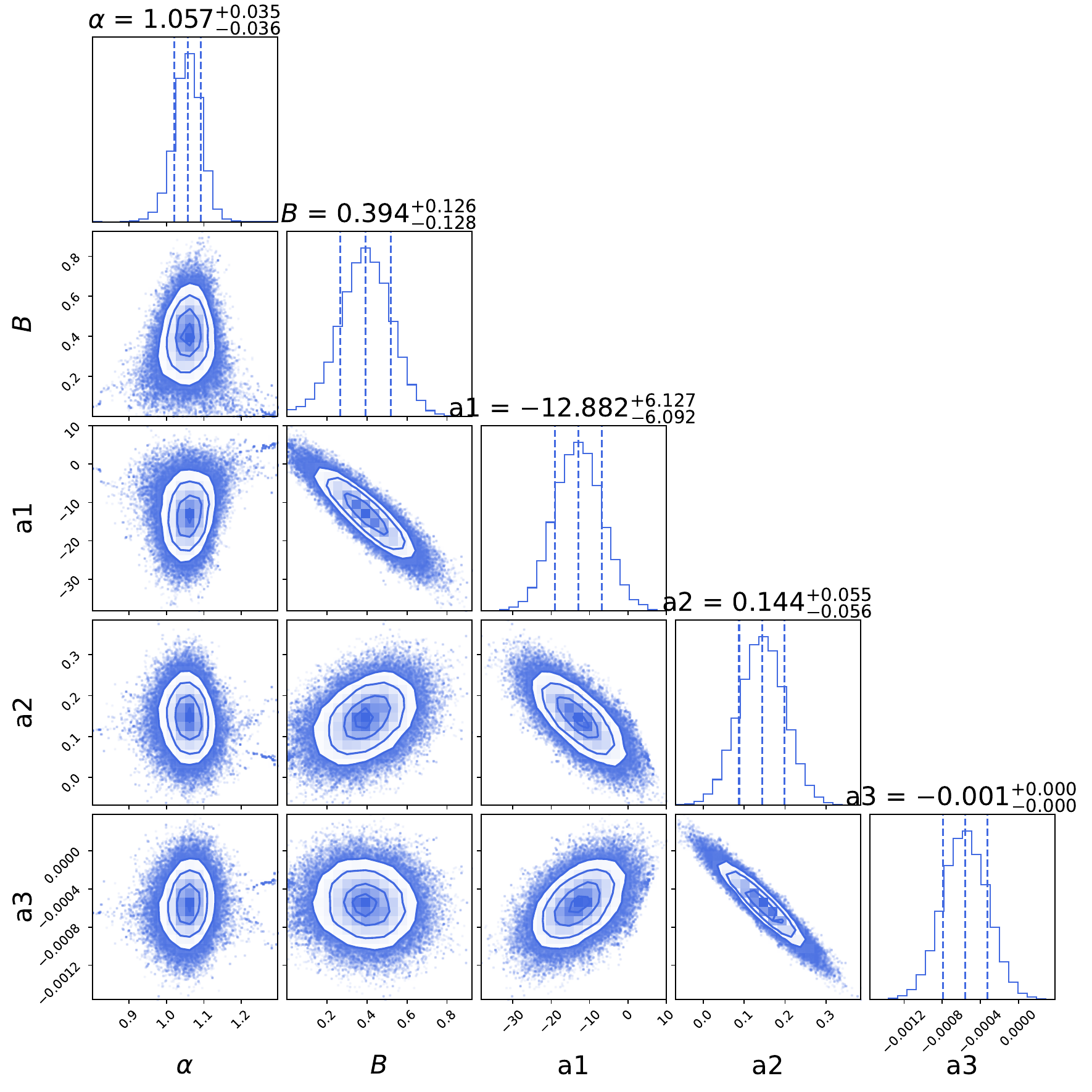}\\
\normalsize{{\textbf{Supplementary Figure 3}}: Constraints from the post-reconstruction GI correlation functions using MCMC sampling with fiducial model. The central values are medians, and the errors are the 16 \& 84
percentiles after other parameters are marginalized over.}\label{fig:mcmc_post}
\end{figure}

\begin{figure}
\centering
\includegraphics[width=1.0\textwidth]{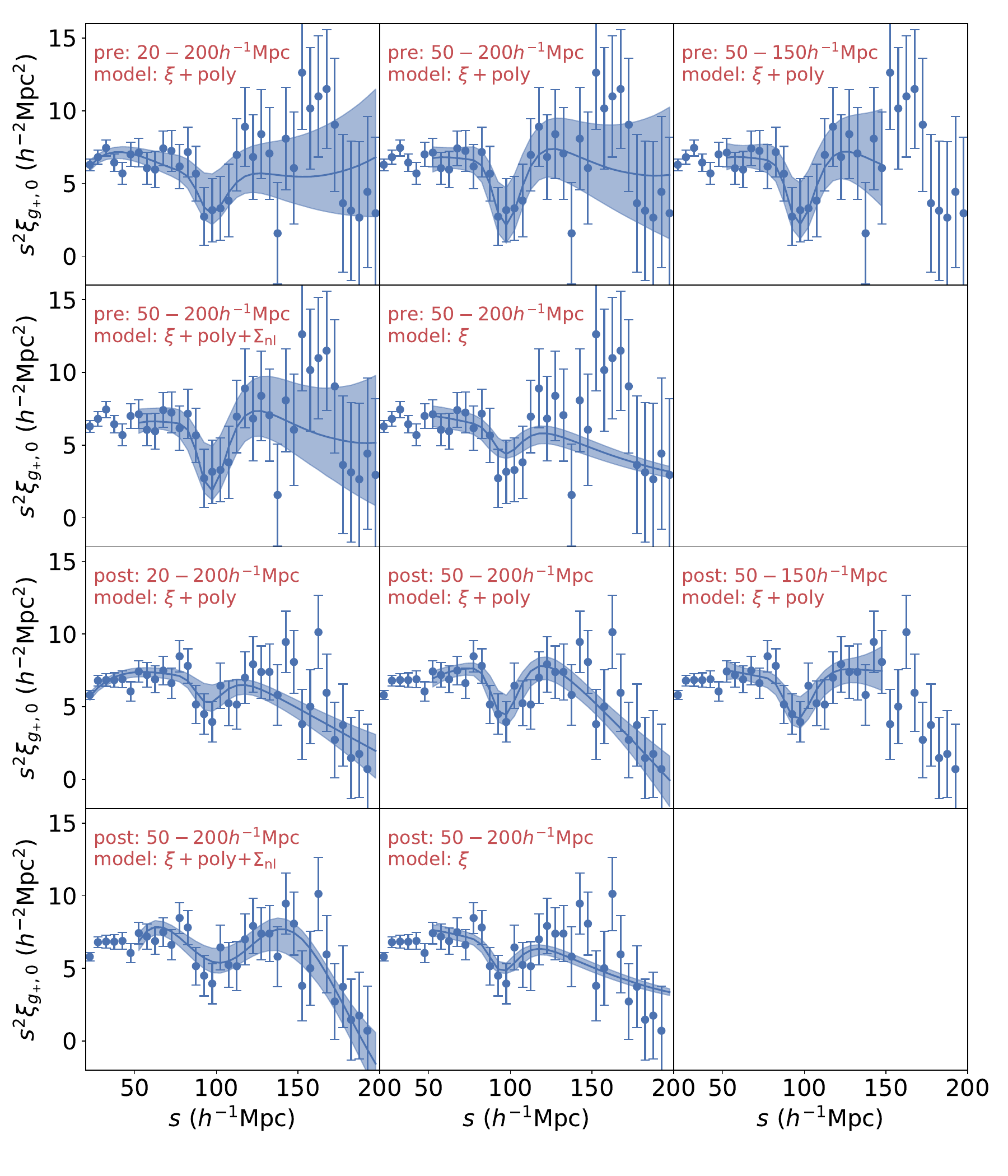}\\
\normalsize{{\textbf{Supplementary Figure 4}}: Modelings of the GI correlation functions with different fitting recipes. Dots with error bars show the mean and standard error of the mean of clustering measurements. Errors are from the diagonal elements of the jackknife covariance matrices estimated using 400 subsamples. Lines with shadows are the best-fit models and $68\%$ confidence level regions derived from the marginalized posterior distributions.}\label{fig:fitting_all}
\end{figure}

\begin{figure}
\centering
\includegraphics[width=1.0\textwidth]{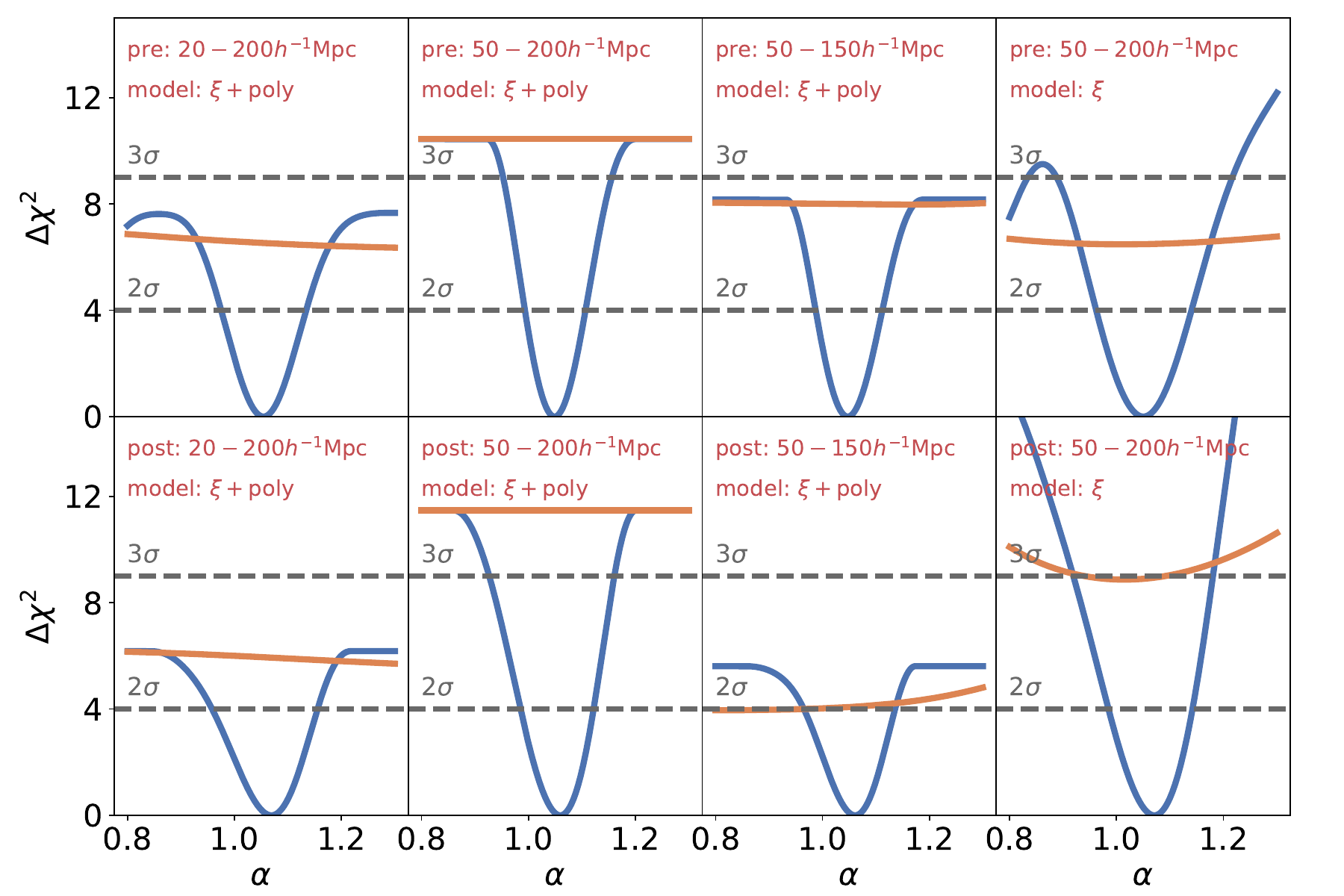}\\
\normalsize{{\textbf{Supplementary Figure 5}}: Plot of $\Delta \chi^2$ versus $\alpha$ for different fitting recipes. Orange lines show the $\Delta \chi^2$ for non-BAO models and blue lines for BAO model.}\label{fig:chi2_all}
\end{figure}

\begin{table}
    \renewcommand{\arraystretch}{0.8}
    \begin{threeparttable}\small
    \begin{tabular}{cccccccc}
         \hline
         \hline
         data & CF & jackknife & model\tnote{a} & range & $\alpha$ & $\Sigma_{{\rm{nl}}}$& $\chi^2/{\rm{d.o.f.}}$ \\
         &&&&($h^{-1}{\rm{Mpc}}$)&&($h^{-1}{\rm{Mpc}}$)\\
         \hline
         \hline
         pre-recon & GI & 400 & $\xi$+poly & [20, 200] & $1.058^{+0.042}_{-0.040}$& 0.0 &1.04\\
         pre-recon & GI & 400 & $\xi$+poly & [50, 200] & $1.050^{+0.030}_{-0.028}$& 0.0 &0.97\\
         pre-recon & GI & 400 & $\xi$+poly & [50, 150] & $1.050^{+0.033}_{-0.031}$& 0.0 &0.75\\
         pre-recon & GI & 400 & $\xi$+poly+$\Sigma_{\rm{nl}}$ & [50, 200] & $1.055^{+0.033}_{-0.032}$& $0.4^{+6.8}_{-6.5}$ &1.01\\
         pre-recon & GI & 400 & $\xi$ & [50,200] & $1.055^{+0.047}_{-0.046}$& 0.0 &1.02\\
         pre-recon & GI & 200 & $\xi$+poly & [50, 200] & $1.047^{+0.032}_{-0.030}$& 0.0 &0.96\\
         pre-recon & GI & 800 & $\xi$+poly & [50, 200] & $1.042^{+0.029}_{-0.029}$& 0.0 &0.98\\
         pre-recon & GI & 1600 & $\xi$+poly & [50, 200] & $1.045^{+0.028}_{-0.031}$& 0.0 &0.96\\
         pre-recon & GI Full\tnote{b}& 400 & $\xi$+poly & [50, 200] & $1.065^{+0.067}_{-0.072}$& 0.0 &0.70\\
         \hline
         post-recon & GI & 400 & $\xi$+poly & [20, 200] & $1.070^{+0.046}_{-0.052}$& 0.0 &1.09\\
         post-recon & GI & 400 & $\xi$+poly & [50, 200] & $1.057^{+0.035}_{-0.036}$& 0.0 &0.87\\
         post-recon & GI & 400 & $\xi$+poly & [50, 150] & $1.057^{+0.041}_{-0.049}$& 0.0 &0.55\\
         post-recon & GI & 400 & $\xi$+poly+$\Sigma_{\rm{nl}}$ & [50, 200] & $1.068^{+0.055}_{-0.053}$& $29.0^{+7.3}_{-9.5}$ &0.90\\
         post-recon & GI & 400 & $\xi$ & [50, 200] & $1.071^{+0.037}_{-0.041}$& 0.0 &1.05\\
         post-recon & GI & 200 & $\xi$+poly & [50, 200] & $1.063^{+0.036}_{-0.040}$& 0.0 &0.84\\
         post-recon & GI & 800 & $\xi$+poly & [50, 200] & $1.047^{+0.033}_{-0.033}$& 0.0 &0.90\\
         post-recon & GI & 1600 & $\xi$+poly & [50, 200] & $1.055^{+0.034}_{-0.036}$& 0.0 &0.89\\
         post-recon & GI Full & 400 & $\xi$+poly & [50, 200] & $1.058^{+0.058}_{-0.078}$& 0.0 &1.15\\
         \hline
         post-recon & GG & 400 & $\xi$+poly & [50, 200] & $0.986^{+0.013}_{-0.013}$& 4.6 &0.92\\
         post-recon & GG+GI & 400 & $\xi$+poly & [50, 200] & $0.996^{+0.012}_{-0.012}$& 4.6+0.0 &0.90\\
         \hline
         \hline
    \end{tabular}
     \begin{tablenotes}
     \footnotesize
     \item[a] 'poly' indicates using polynomial to fit the broad band shape and '$\Sigma_{{\rm{nl}}}$' means setting $\Sigma_{{\rm{nl}}}$ as a free parameter.
     \item[b] Using the whole CMASS sample as the tracers of the ellipticity field without $n$ and $q$ selections.
     \end{tablenotes}
     \end{threeparttable}
    \label{tab:t1}
    \vspace{0.3 cm}
\normalsize    
{{\textbf{Supplementary Table 1}}: Constrains of $\alpha$ from different correlation functions, jackknife resampling scales, models and fitting ranges.}
\end{table}

In {\textbf{Supplementary Figure 4}} and {\textbf{5}} and {\textbf{Supplementary Table 1}}, we show the constrains of $\alpha$ from different correlation functions, jackknife resampling scales, models and fitting ranges. Comparing different values of $N_{\rm{JK}}$, the results are relatively stable when $N_{\rm{JK}}\geqslant400$, verifying that $N_{\rm{JK}}=400$ is a reasonable choice. The fitting results are slightly affected by the models and fitting ranges, but all these fitting recipes results in a $2\sim3 \sigma$ detection of BAO and $3\sim5\%$ constraints on $\alpha$ for both pre- and post-reconstruction GI measurements, which proves that our BAO detection and distance constraints are relatively robust to the details of modeling. Results with the whole CMASS sample as the tracers of the ellipticity field are also included in {\textbf{Supplementary Table 1}}, confirming that our sample selection can improve the precision of GI BAO measurements.

\end{document}